\documentclass[10pt,twocolumn]{IEEEtran} 
\usepackage{fancyhdr}
\usepackage{epsfig}
\usepackage{threeparttable}
\usepackage{epsf,epsfig}
\usepackage{amsthm}
\usepackage{amsmath}
\usepackage{amssymb}
\usepackage{amsfonts}
\usepackage[noadjust]{cite}
\usepackage{dsfont}
\usepackage{subfigure}
\usepackage{color}
\usepackage{enumerate}

\newtheorem{lemma}{Lemma}
\newtheorem{corollary}{Corollary}

\newtheorem{proposition}{Proposition}

\def\E{\mathsf{E}}
\def\SINR{\mathsf{SINR}}
\def\SNR{\mathsf{SNR}}
\def\SIR{\mathsf{SIR}}

\def\l{\left}
\def\r{\right}
\def\({\left(}
\def\){\right)}

\def\[{\left[}
\def\]{\right]}

\def\papertitle{\huge Resource Management and Cell Planning in Millimeter-Wave Overlaid Ultra-Dense Cellular Networks}

\IEEEoverridecommandlockouts 

\begin{document}

\title{ \fontsize{23}{24}\selectfont \papertitle}


\author{Jihong Park, Seong-Lyun Kim, and Jens Zander
\thanks{J. Park and S.-L. Kim are with Dept. of Electrical \& Electronic Engineering, Yonsei University, Seoul, Korea (email: jhpark.james@yonsei.ac.kr, slkim@yonsei.ac.kr).  }
\thanks{J. Zander is with Wireless@KTH, KTH -- The Royal Institute of Technology, Stockholm, Sweden (email: jensz@kth.se).}}


\maketitle

\begin{abstract} 
This paper proposes a cellular network exploiting millimeter-wave (mmWave) and ultra-densified base stations (BSs) to achieve the far-reaching 5G aim in downlink average rate. The mmWave overlaid network however incurs a pitfall that its ample data rate is only applicable for downlink transmissions due to the implementation difficulty at mobile users, leading to an immense difference between uplink and downlink rates. We therefore turn our attention not only to maximize downlink rate but also to ensure the minimum uplink rate. With this end, we firstly derive the mmWave overlaid ultra-dense cellular network spectral efficiencies for both uplink and downlink cases in closed forms by using stochastic geometry via a lower bound approximation. In a practical scenario, such tractable results of the proposed network reveal that incumbent micro-wave ($\mu$Wave) cellular resource should be mostly dedicated to uplink transmissions in order to correspond with the mmWave downlink rate improvement. Furthermore, increasing uplink rate via $\mu$Wave BS densification cannot solely cope with the mmWave downlink/uplink rate asymmetry, and thus requires additional $\mu$Wave spectrum in 5G cellular networks.
\end{abstract}

\section{Introduction}

The scarcity of cellular frequency for achieving the 1,000x higher data rate in 5G brings about exploiting the millimeter-wave (mmWave) frequency band having hundreds times more spectrum amount \cite{Rappaport5G:13}. This approach however has two major drawbacks. Firstly, mmWave signals are vulnerable to physical blockages, yielding severe distance attenuation. Secondly, the use of extremely wide mmWave bandwidth makes uplink mmWave transmissions at mobiles demanding due to high peak-to-average-ratio (PAPR) \cite{SamsungmmWave:11}, leading to significant rate difference between uplink and downlink.

To compensate both drawbacks, we consider base station (BS) ultra-densification. It is another promising way to enhance the data rate by increasing per-user available resource amount \cite{Andrews5G:14} and even by improving  per-user average spectral efficiency \cite{Holistic13,JHPark:14}. This study combines such complementary two methods, and thereby proposes a mmWave overlaid ultra-dense cellular network.

The proposed network in distinction from other existing heterogeneous networks is the asymmetric uplink and downlink structure where mmWave band only operates for downlink transmissions (see Fig. 1). The reason is even the state-of-the-art mmWave supporting power amplifiers are low energy efficient, so uplink mmWave transmissions are demanding of mobile users \cite{SamsungmmWave:11}. Uplink communication therefore solely resorts to the current micro-wave ($\mu$Wave) cellular frequency band. Due to the scarcity of the $\mu$Wave spectrum, the uplink data rate cannot keep up with the rising ample downlink data rate as mmWave resource and BS densification grow, which may hinder consistent user experiences. This uplink/downlink asymmetry motivates to turn our attention toward assuring the minimum uplink average rate, and engenders the mmWave overlaid cellular network design problem: \emph{how to maximize the downlink average rate while guaranteeing a target uplink average rate}. 

We answer this question from the radio resource management and cell planning perspectives. Exploiting stochastic geometry and the technique proposed in our preliminary work \cite{JHPark:14}, we derive downlink mmWave and uplink/downlink $\mu$Wave spectral efficiencies in closed forms. Utilizing these tractable results, we analyze the impacts of resource management and cell planning on downlink and uplink average rates. Such results provide the design guidelines of the mmWave overlaid ultra-dense cellular networks. 

The main contributions of this paper are listed as follows.
	\begin{enumerate}[1.]
	\item \emph{Most of the $\mu$Wave resource should be dedicated to uplink transmissions} in order to guarantee the uplink average rate by at least $3\%$ of the downlink rate in a practical scenario where $\mu$Wave and mmWave resource amounts respectively are $20$ MHz and $500$ MHz \cite{SamsungGC:13} (see Proposition 3 and Fig. 6). This runs counter to the current resource allocation trend that is likely to allocate more resource to downlink transmissions \cite{Rel12Beyond:13}.
	\item  To achieve the ever-growing downlink average rate while guaranteeing uplink average rate, \emph{densifying $\mu$Wave BS cannot be a sole remedy in practice} (see Proposition 4), \emph{but should be in conjunction with procuring more $\mu$Wave spectrums} (see Corollary 2). The reason behind is more spectrum amount linearly increases the uplink average rate while BS densification logarithmically increases the rate (see Proposition 1).
	\item The spectral efficiencies in uplink/downlink $\mu$Wave (see Proposition 1) and downlink mmWave bands (see Proposition 2) under an ultra-dense environment are derived in closed forms via a lower bound approximation, which reveals \emph{BS densification logarithmically increases the spectral efficiency}. 
	\end{enumerate}

Due to the lack of the space, the omitted proofs of propositions and lemmas are deferred to: \emph{http://tiny.cc/jhparkgc15pf}.



\begin{figure}
\centering
 	\includegraphics[width=9cm]{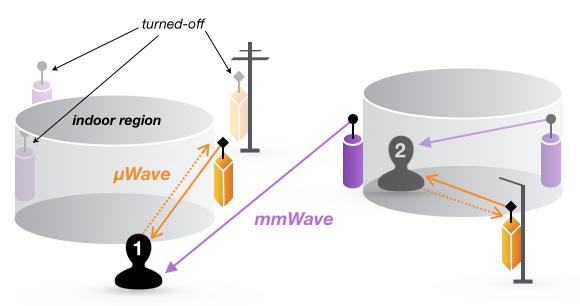}  
	\caption{Illustration of a mmWave overlaid cellular network where downlink signals are transmitted via both mmWave and $\mu$Wave bands whereas the uplink only via $\mu$Wave band. Indoor region is not penetrated by mmWave but $\mu$Wave signals. Users associate with the nearest non-blocked mmWave BSs (outdoor user 1 associates with a farther mmWave BS than the nearest indoor BS) as well as with the nearest $\mu$Wave BSs without any restriction (indoor user 2 associates with the outside nearest $\mu$Wave BS). The BSs having no serving users are turned-off.}
	\label{Fig:NetULDL}
\end{figure}

\section{System Model} \label{Sect:SysModel}

\subsection{Milimeter-Wave Overlaid Cellular Network}
The proposed network comprises: (i) mmWave BSs whose locations follow a two-dimensional homogeneous Poisson point process (PPP) $\Phi_\text{m}$ with density $\lambda_\text{m}$; and (ii) $\mu$Wave BSs whose coordinates follow a homogeneous PPP $\Phi_\mu$ with density $\lambda_\mu$, independent of $\Phi_\text{m}$. Due to the implementational difficulty of mmWave transmissions at mobile users, mmWave BS only supports downlink mode whereas $\mu$Wave BS provides both downlink and uplink modes. Uplink transmissions are therefore resort to solely depend on $\mu$Wave BSs. A BS having no serving user is turned off (see the transparent BSs in Fig. 1).

Mobile user coordinates independently follow a homogeneous PPP $\Phi_u$ with density $\lambda_\text{u}$. Without loss of generality, $\Phi_u$ represents both downlink and uplink users. Users receive downlink signals via both mmWave and $\mu$Wave simultaneously \cite{Rel12Beyond:13} while transmitting uplink signals only via $\mu$Wave. Specifically, downlink users associate with their nearest non-blocked mmWave BSs, and also independently with their nearest $\mu$Wave BSs as in \cite{Rel12Beyond:13,Ericsson:12}. Uplink users associate with the nearest $\mu$Wave BSs that are identical with their $\mu$Wave downlink associated BSs but the mmWave associated BSs. Such associations are visually shown by Fig. 1.

\subsection{Indoor/Outdoor Model} \label{Sect:InOut}
Consider indoor regions whose boundaries are mmWave impenetrable walls. Followed by a Boolean model \cite{HallBook:CoverageProcess:1988}, the indoor regions are regarded as uniformly distributed circles having radius $R$ with density $\lambda_g$. For simplicity without loss of generality, assume the indoor regions always guarantee line-of-sight (LOS) communications. Additionally, we neglect the overlapping indoor regions that can be compensated by a sufficiently large network. Note that $\mu$Wave signals are not affected by the indoor walls thanks to their high diffraction and penetration characteristics. The indoor complementary regions are outdoor regions.

\subsection{Channel Model} \label{Sect:Channel}
Our channel model is three-fold in order to capture the different propagation behaviors of outdoor/indoor mmWave and $\mu$Wave signals. 
\subsubsection{Outdoor mmWave Channel} 
A mmWave antenna array directionally transmits a downlink signal with unity power to its associated user, and the signal experiences path loss attenuation with the exponent $\alpha_\text{m}>2$ as well as Rayleigh fading with unity mean. The transmitted directional beam has the main lobe angle $\theta$ (radian), and the received signal powers at the same distances within $\theta$ are assumed to be identical.

Users are able to receive mmWave signals only if there exist no indoor walls along the paths to their associated BSs. To specify this event, consider a typical user $U_0$ located at the origin and an arbitrary BS at $X_i$ with distance $r_i$. Let $\mathbf{u_i}$ denote the opposite unit direction vector of the signal transmission direction  from $X_i$, defined as $|X_i|/r_i$. Define $L(\mathbf{u_i})$ as the non-blockage distance indicating the line length from $U_0$ with the direction $\mathbf{u_i}$ to the point when the line firstly intersects an impenetrable indoor wall. A user then can receive a transmitted signal if the condition $L(\mathbf{u_i})>r_i$ holds. 

For interfering links, we consider the interferers as the undesired active BSs whose serving directions, the main lobe centers, pointing to $U_0$. It leads to $\Theta_i$ antenna gain, which is an increasing function of $\theta$ as well as the number of the BS's serving users. At $U_0$ when located outside the indoor regions, the corresponding mmWave signal-to-interference-plus-noise ($\SINR$) is represented  as:
\begin{align}
\small \SINR_{\text{m.out}} := \left\{\begin{array}{lll}
             \frac{ \eta r^{-\alpha_\text{m}}  }{\sum_{i\in \Phi_\text{m.out}} \Theta_i \eta_i {r_i}^{-\alpha_\text{m}}+ \sigma^2} & \text{if }L(\mathbf{u})>r \\
              0 & \text{otherwise}
            \end{array}\right.
\end{align}\normalsize
where $\Phi_{\text{m.out}}$ indicates active non-blocked mmWave interfering outdoor BSs, $\eta_i$ fading power, and $\sigma^2$ noise power.

\subsubsection{Indoor mmWave Channel}
Path loss exponent is set as $2$ since there is no mmWave blockages within indoor regions as described in Section \ref{Sect:InOut}. The rest of the settings are the same as in the case of the outdoor mmWave. At $U_0$ when located within indoor regions, the mmWave $\SINR$ is given as:
\small\begin{equation}
\SINR_{\text{m.in}} := \frac{  \eta r^{-2}  }{\sum_{i\in \Phi_\text{m.in}} \Theta_i \eta_i {r_i}^{-2}+ \sigma^2}
\end{equation}\normalsize
where $\Phi_{\text{m.in}} $ denotes active mmWave interfering indoor BSs.

\subsubsection{$\mu$Wave Channel}
A transmitted signal with unity power experiences path loss attenuation with the exponent $\alpha_\mu>2$ as well as Rayleigh fading, resulting in the fading gain $g$ that follows an exponential distribution with mean $1$. At $U_0$, the $\mu$Wave $\SINR$ is given as:
\small\begin{align}
\SINR_{\mu} &:= \frac{g r^{-\alpha_\mu}}{\sum_{i\in \Phi_{\mu}} g_i {r_i}^{-\alpha_\mu}+ \sigma^2}
\end{align}\normalsize
where $\Phi_{\mu} $ represents active $\mu$Wave interfering BSs.

\section{Spectral Efficiency in Ultra-Dense Cellular Networks} \label{Sect:RateUDN}

This section derives closed-form mmWave and $\mu$Wave spectral efficiencies, defined as ergodic capacity $\E \log[1 + \SINR]$ in units of nats/sec/Hz (1bit $\approx$ 0.693 nats), to be utilized for resource allocation in Section \ref{Sect:RscCell}. For brevity, we consider a downlink network in this section unless otherwise noted.

\subsection{Ultra-Densification Effect}
Throughout this study, an ultra-dense environment is of our interest. The ultra-densification implies BS density is much larger than user density such that the distance to an interfering BS at a typical receiver can be approximated by the distance to the interfering BS's nearest user. 

To be more rigorous, define active probability $p_a$ that an arbitrary $\mu$Wave BS is turned-on, i.e. serving at least a single user within its coverage. According to Proposition 1 in \cite{Yu2011},
\begin{eqnarray}
p_a = \l\{1-(1 + 3.5^{-1}\lambda_\text{u} / \lambda_\mu)^{-3.5}\r\} \overset{(a)}{\approx} \lambda_\text{u}/\lambda_\mu \label{Eq:ActiveProb}
\end{eqnarray}
where $(a)$ follows from Taylor expansion for $\lambda_\mu \gg \lambda_\text{u}$.

This tendency in ultra-dense networks results in the active BS density $p_a \lambda_\mu$ converging to $\lambda_\text{u}$. The result implies ultra-densification enables to increase the received signal power without incurring any interference increment since interference is delimited by user density independent of the BS densification. The same tendency applies to the mmWave BS active probability, so $p_a$ henceforth denotes both $\mu$Wave and mmWave active probabilities.

\begin{figure}
\centering
\includegraphics[width=9cm]{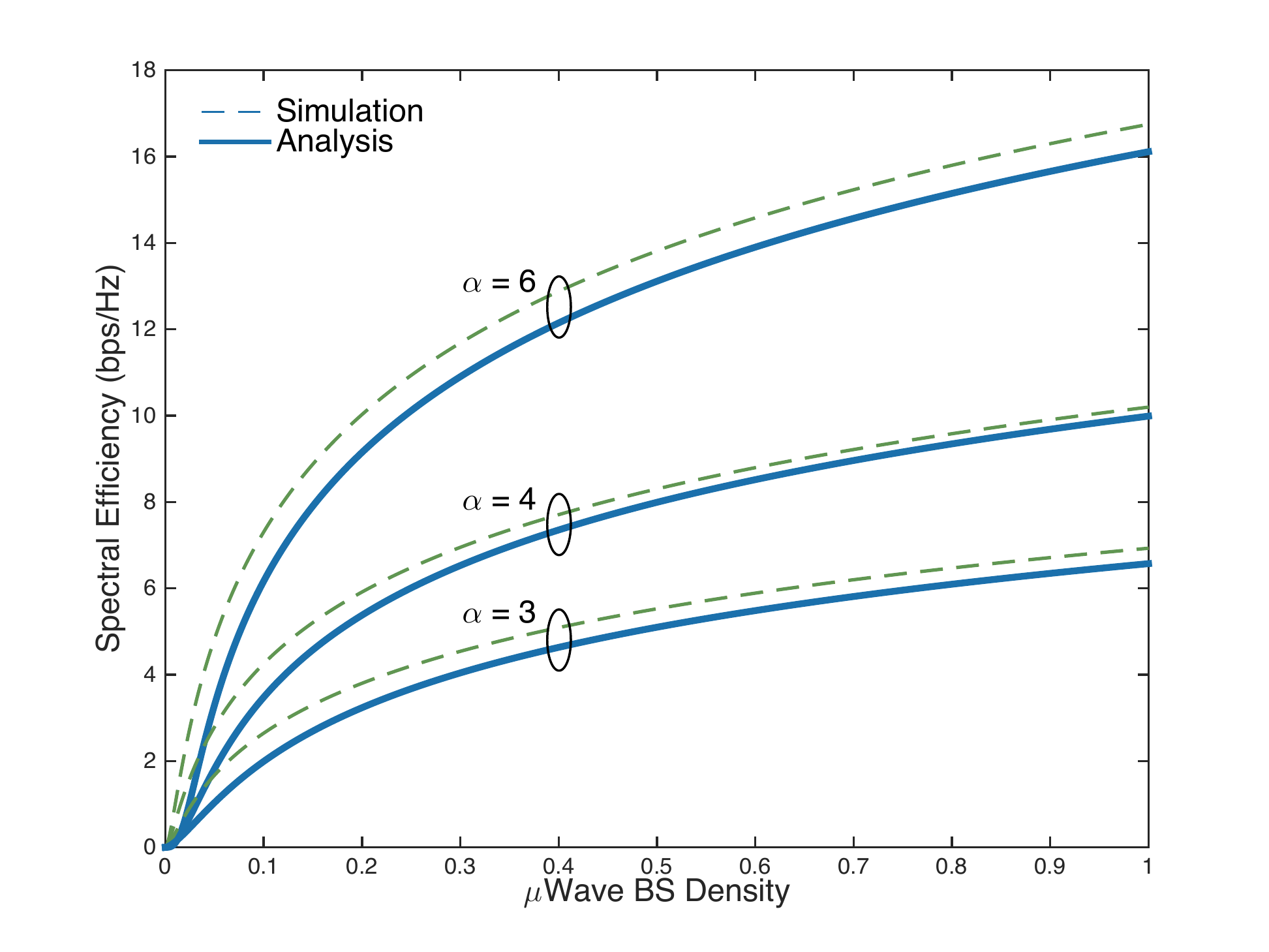}  
	\caption{ Spectral efficiency of a typical downlink (or uplink) $\mu$Wave user for path loss exponents $\alpha=3$, $4$, and $6$ where user density $\lambda_\text{u}$ is set as $0.02$. Compared to the simulation curve when $\alpha=4$, the analytic curve achieves: 81.75\%, 90.9\%, and 97.96\% for $\lambda_\mu=0.1$, $0.2$, and $1$ respectively.}
	\label{Fig:CapApprox}
\end{figure}

\subsection{$\mu$Wave Spectral Efficiency} \label{Sect:muWaveSE}
This section derives the tractable but accurate lower bounds of uplink and downlink $\mu$Wave spectral efficiencies. Firstly, consider uplink spectral efficiency that is deemed difficult to be analyzed in a stochastic geometric point of view. 

To specify the difficulty, we compare downlink and uplink scenarios. In a downlink case at $U_0$, all interfering BSs are farther than the associated BS. In an uplink case at a typical BS $X_0$, on the other hand, interfering users may be located closer than the associated user $U_0$ since the nearest association is not determined by $X_0$ but users. This distance disorder of uplink interferers and $U_0$ makes the uplink aggregate interference analysis complicated. 

Ultra-dense environment however allows to detour the distance disorder problem by enabling an uplink-downlink reciprocity thanks to short user-to-active BS distances, which leads to the following Lemma.

\begin{lemma}\emph{
\emph{(Uplink--Downlink Reciprocity)} For $\lambda_\mu \gg \lambda_\text{u}$, uplink and downlink $\mu$Wave spectral efficiencies are identical.\\
}\end{lemma}

Secondly, consider downlink $\mu$Wave spectral efficiency. Its closed-form result has been derived in our preliminary work \cite{JHPark:14}, but is restated here for the integrity of this study in the following Proposition.

\begin{proposition}\emph{
\emph{(Uplink/Downlink $\mu$Wave)} At $U_0$ for $\lambda_\mu \gg \lambda_\text{u}$, uplink (or downlink) $\mu$Wave spectral efficiency $\gamma_\mu$ is lower bounded as follows.
\begin{equation}
\gamma_{\mu} >  \log\( 1 + \[ \frac{\lambda_\mu}{\rho_\mu \lambda_\text{u}}\]^{\alpha_\mu/2}\)
\end{equation}
where $\rho_\mu := \int_{0}^\infty 1/(1 + u^{\alpha_\mu /2}) du$ \\
}\end{proposition}
The result indicates that increasing BS density logarithmically improves the $\mu$Wave spectral efficiency. The tightness of the proposed lower bound is numerically verified by Fig. 2. The figure also illustrates increasing path loss exponent enhances the spectral efficiency since its resultant interference reduction dominates the desired signal attenuation under an interference-limited regime as expected in \cite{Yu2011}.

\begin{figure}
\centering
\includegraphics[width=9cm]{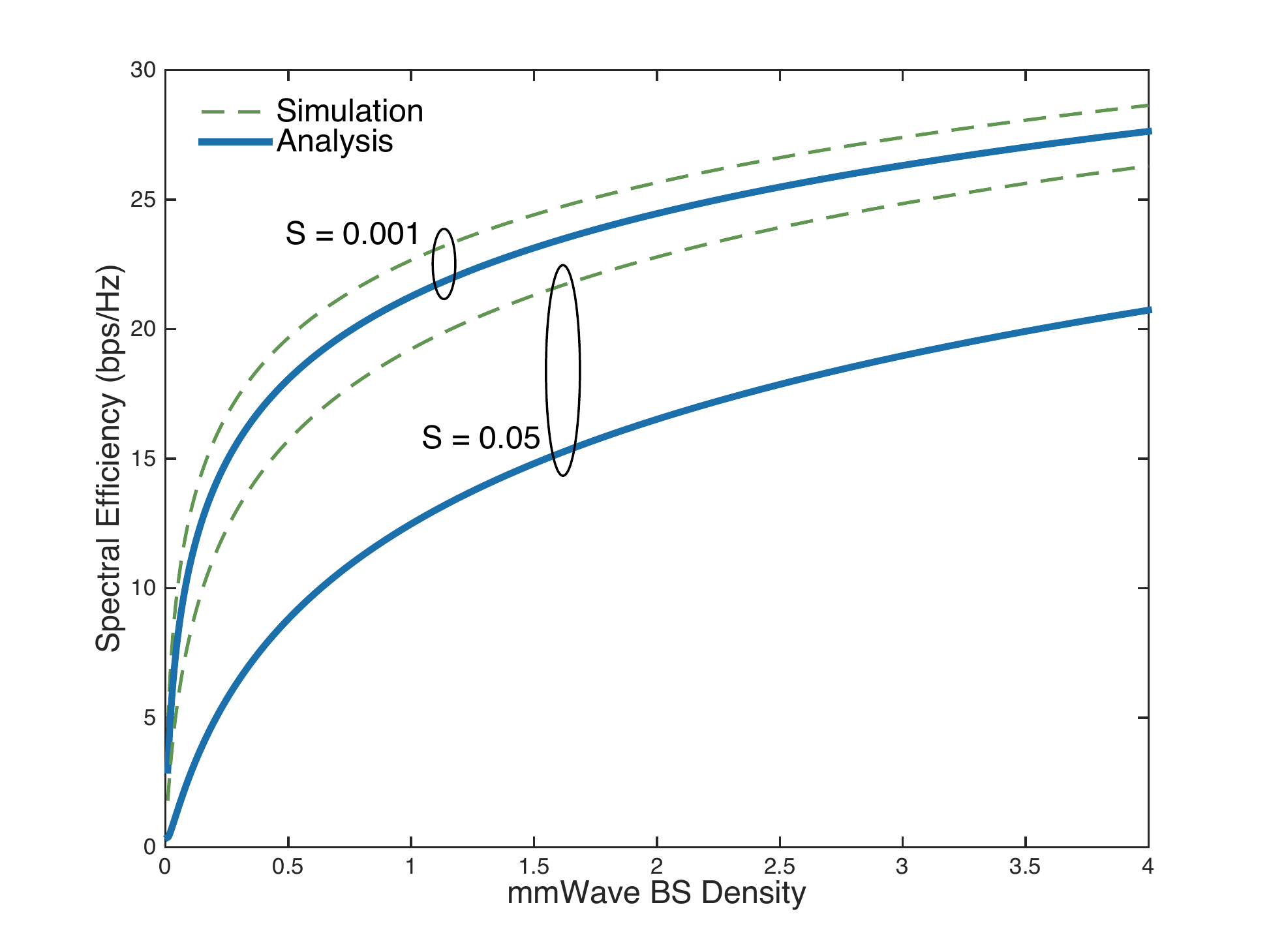}  
	\caption{Spectral efficiency of a typical \emph{outdoor} downlink mmWave user for the outdoor path loss exponent $\alpha_\text{m}=5.76$ \cite{Rappaport5G:13} where user density $\lambda_\text{u} = 0.02$, indoor region density $\lambda_g = 0.1$, and main lobe beam width $10^\circ$.}
	\label{Fig:SEmm_out}
\end{figure}

\subsection{mmWave Spectral Efficiency} \label{Sect:SEmmWave}
This section derives the downlink mmWave spectral efficiencies under outdoor and indoor environments. A notable mmWave characteristic in distinction from the preceding $\mu$Wave analysis is indoor wall impenetrability. Its impact on the mmWave spectral efficiency is captured by a single indoor region area $S$ defined as $\pi R^2$. We confine to considering a large outdoor region compared to indoor region, i.e. small $S$. Specifically, $S$ is sufficiently small in such a way that each indoor region contains at most a single user, leading to a noise-limited indoor environment and conversely an interference-limited outdoor environment. 

Firstly, in an interference-limited outdoor region, mmWave spectral efficiency $\gamma_\text{m.out}$ is analytically represented in the following Lemma.


\begin{lemma}\emph{\emph{(Downlink Outdoor mmWave)} At $U_0$ when located outside indoor regions for $\lambda_\text{m} \gg \lambda_\text{u}$, downlink outdoor mmWave spectral efficiency $\gamma_{\text{m.out}}$ is lower bounded as follows.
\begin{equation}
\gamma_{\text{m.out}} >   \log\( 1 + \frac{2 \pi }{\theta}\[  \frac{ e^{\lambda_g S}\lambda_\text{m} }{ \rho_\text{m}  \lambda_\text{u}} \]^{\frac{\alpha_\text{m}}{2}} \)^{1 -  \sqrt{\frac{S}{\lambda_\text{m}}} } \label{Eq:Lemma3}
\end{equation}
where $\rho_\text{m} := \int_{0}^\infty 1/(1 + u^{\alpha_\text{m} /2}) du$ \\
}\end{lemma}
\vspace{-10pt}The result implies BS densification not only improves $\SIR$, but also overcomes indoor wall blockages (see the logarithm function's exponent in \eqref{Eq:Lemma3}), leading to the spectral efficiency enhancement. Increasing indoor area $S$ decreases outdoor interference while increasing desired signal blockage effect. The latter can be fully compensated by BS densification, so larger $S$ ensures better outdoor spectral efficiency under an ultra-dense environment. The tightness of the analytic result is numerically validated, and visualized in Fig. 3.

\begin{figure}
\centering
\includegraphics[width=9cm]{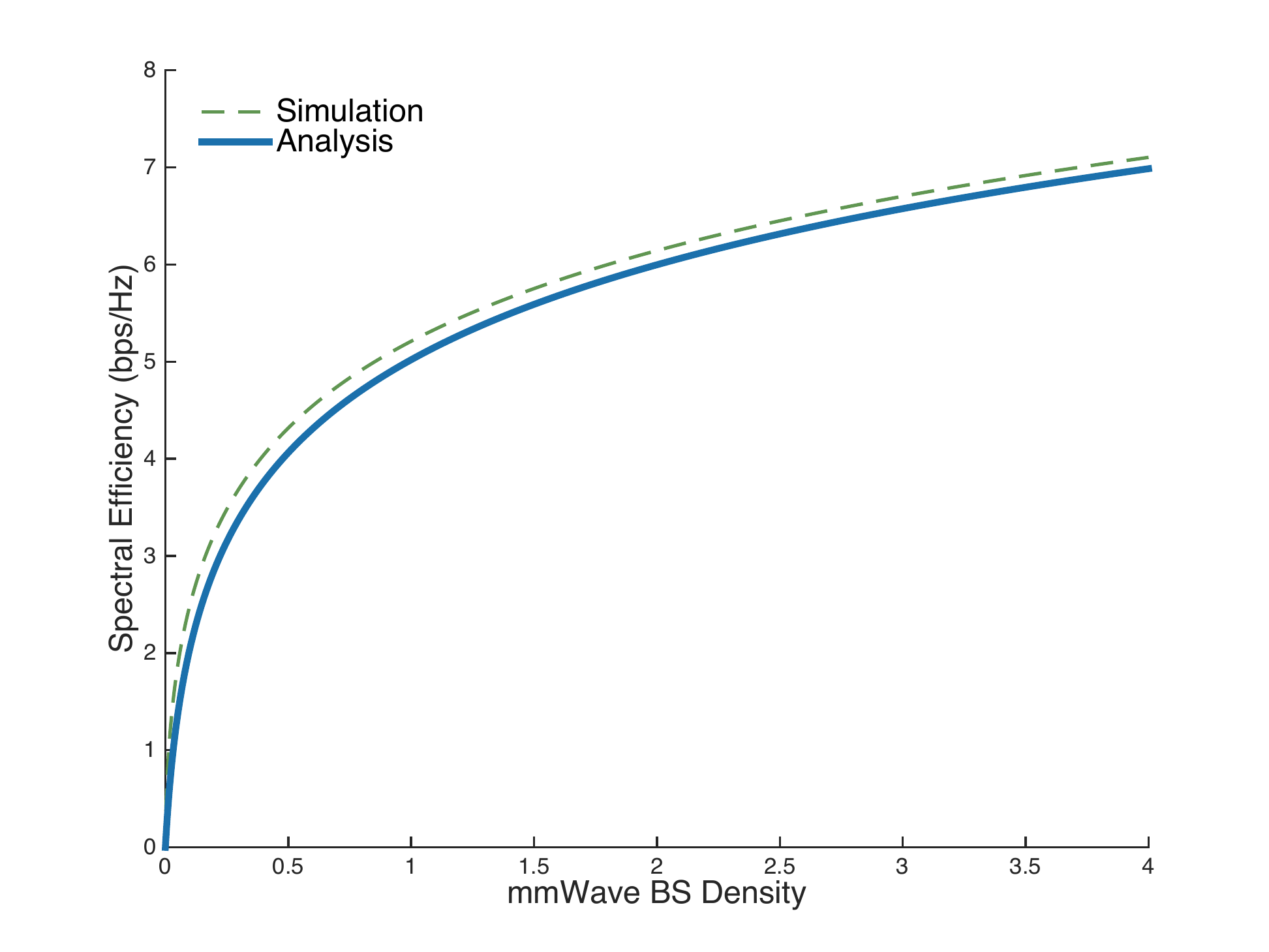}  
	\caption{Spectral efficiency of a typical \emph{indoor} downlink mmWave user for the indoor path loss exponent is set as $2$, and user density $\lambda_\text{u}=0.02$.}
	\label{Fig:SEmm_in}
\end{figure}

Secondly, in a noise-limited indoor region, mmWave spectral efficiency $\gamma_\text{m.in}$ is given in the following Proposition.
\begin{lemma}\emph{\emph{(Downlink Indoor mmWave)} At $U_0$ when located within indoor regions for $\lambda_\text{m} \gg \lambda_\text{u}$, downlink indoor mmWave spectral efficiency $\gamma_\text{m.in}$ is lower bounded as follows.
\begin{equation}
\gamma_{\text{m.in}} >    \log\( 1 + \frac{\pi  \lambda_\text{m}}{\sigma^2} \)
\end{equation}\\
}\end{lemma}\vspace{-25pt}
The relationship between BS density $\lambda_\text{m}$ and received power intuitively interprets the result above. The received power scales with $r^{-2}$ for indoor mmWave path loss exponent $2$ while $\E[r]$ in \cite{StoyanBook:StochasticGeometry:1995} scales with $1/\sqrt{\lambda_\text{m}}$, resulting in the linear scaling of $\SNR$ with $\lambda_\text{m}$. The tightness of the analysis is numerically verified as shown in Fig. 4.

Combining the outdoor and indoor mmWave spectral efficiencies (Lemmas 3 and 4), the following result provides the overall downlink mmWave spectral efficiency $\gamma_\text{m}$ as below.


\begin{proposition}\emph{\emph{(Downlink mmWave)} At $U_0$ for $\lambda_\text{m} \gg \lambda_\text{u}$, downlink mmWave spectral efficiency $\gamma_\text{m}$ at $U_0$ is lower bounded as follows.
\begin{equation}\small
\gamma_\text{m} > \log\(1 +  \frac{\pi {\lambda_\text{m}}^{ \(\frac{\alpha_\text{m}}{2}-1 \)e^{-\lambda_g S} + 1}}{\sigma^2} \[ \frac{2\sigma^2}{\theta} \( \frac{e^{\lambda_g S}}{\rho_\text{m}\lambda_\text{u}} \)^{\frac{\alpha_\text{m}}{2}} \]^{e^{-\lambda_g S}} \) \label{Eq:Prop2}
\end{equation}\normalsize \vspace{-10pt}
}\end{proposition}
The result indicates mmWave downlink spectral efficiency is a logarithmic function of BS density $\lambda_\text{m}$. The exponent of $\lambda_\text{m}$ shows densification is more effective when 1) outdoor mmWave attenuation is severe (large $\alpha_\text{m}$) and/or 2) users are more likely to be in outdoor region (small $\lambda_g S$). In addition, sharper beam (small $\theta$) increases the spectral efficiency. On the other hand, the spectral efficiency decreases with user density since more users bring about larger interference.

%


\section{Resource Management and Cell Planning in Milimeter-Wave Overlaid Ultra-Dense Cellular Networks} \label{Sect:RscCell}
This section analyzes $\mu$Wave resource allocation behaviors, and consequently provides the mmWave overlaid cellular network design guidelines in the perspectives of resource allocation and cell planning.

\vspace{-5pt}

\subsection{Downlink Average Rate with Minimum Uplink Rate Requirement}
Define downlink and uplink average rates $R_\text{d}$ and $R_\text{u}$ as follows.
\begin{eqnarray}
	R_{\text{d}} &:=& W_{\mu.\text{d}} \gamma_\mu + W_\text{m} \gamma_m \label{Eq:RateDL} \\
	R_{\text{u}} &:=& W_{\mu.\text{u}}\gamma_{\mu}
	\end{eqnarray}

Let $T\leq 1$ denote the minimum ratio of uplink to downlink rates. We consider the following problem:
\begin{align}
      \textsf{P1}.&
   \begin{aligned}[t]
    & \underset{W_\mu  }{\text{max}} \; R_{\text{d}}  \notag\\
   \end{aligned}   \notag\\
   &\text{subject to} \notag \\
   & \quad   R_{\text{u}}/R_{\text{d}}   \geq  T  \label{Eq:UplinkQoS} \\
   &\quad  W_{\mu.\text{d}} + W_{\mu.\text{u}} = W
\end{align} 
where $W_{\mu.\text{d}}$, $W_{\mu.\text{u}}$, $W$ respectively denote downlink, uplink, and entire bandwidths. The objective function is maximized when the equality in \eqref{Eq:UplinkQoS} holds, leading to the following $\mu$Wave downlink resource allocation.

\begin{proposition}\emph{\emph{($\mu$Wave Resource Allocation)} If $W\geq T W_\text{m} \gamma_\text{m}/\gamma_\mu$, the following $\mu$Wave downlink resource allocation maximizes the average rate while guaranteeing the minimum uplink rate:
\small\begin{align}
W_{\mu.\text{u}}^* &= \frac{T}{1 + T} \( W + \frac{W_\text{m} \gamma_\text{m}}{\gamma_{\mu}}\)  \quad \text{and } \nonumber\\
W_{\mu.\text{d}}^* &= \frac{1}{1 + T} \( W - \frac{T W_\text{m} \gamma_\text{m}}{\gamma_{\mu}}\); \label{Eq:OptW}
\end{align} \normalsize
where $0 \leq W_{\mu.\text{d}}^*, W_{\mu.\text{u}}^* \leq W$; otherwise, the minimum uplink rate requirement cannot be satisfied.
}\end{proposition}
Increasing mmWave downlink rate ($\lambda_\text{m}$ and/or $W_\text{m}$) leads to less downlink $\mu$Wave allocation $W_{\mu.\text{d}}^*$ and more uplink $\mu$Wave allocation $W_{\mu.\text{u}}^*$. The reason is because the mmWave band provides most of the downlink transmissions without the aid of $\mu$Wave band, and thus all the $\mu$Wave band becomes dedicated to the uplink transmissions to assure the minimum uplink rate requirement. In addition, increasing $\mu$Wave resource $W$ allows more $W_{\mu.\text{d}}^*$ and $W_{\mu.\text{u}}^*$, but the latter increases faster (notice $1/(1+T)\geq T/(1+T)$ for $T\leq 1$). Such uplink biased $\mu$Wave resource allocation tendency is the opposite way of the current resource management trend seeking more downlink resources, to be further elucidated under practical scenarios in Section V. 

Increasing $\mu$Wave BS density $\lambda_\mu$, on the other hand, makes the resource more prone to be allocated to downlink transmissions. Recalling $\mu$Wave uplink-downlink reciprocity in Lemma 2, increasing $\lambda_\mu$ identically improves both uplink and downlink rates. In spite of such identical increments, the uplink/downlink ratio increases since the uplink rate is no larger than downlink rate. This results in increasing $W_{\mu.\text{d}}$ until the equality in \eqref{Eq:UplinkQoS} holds.

\begin{figure}
\centering
 	\includegraphics[width=9cm]{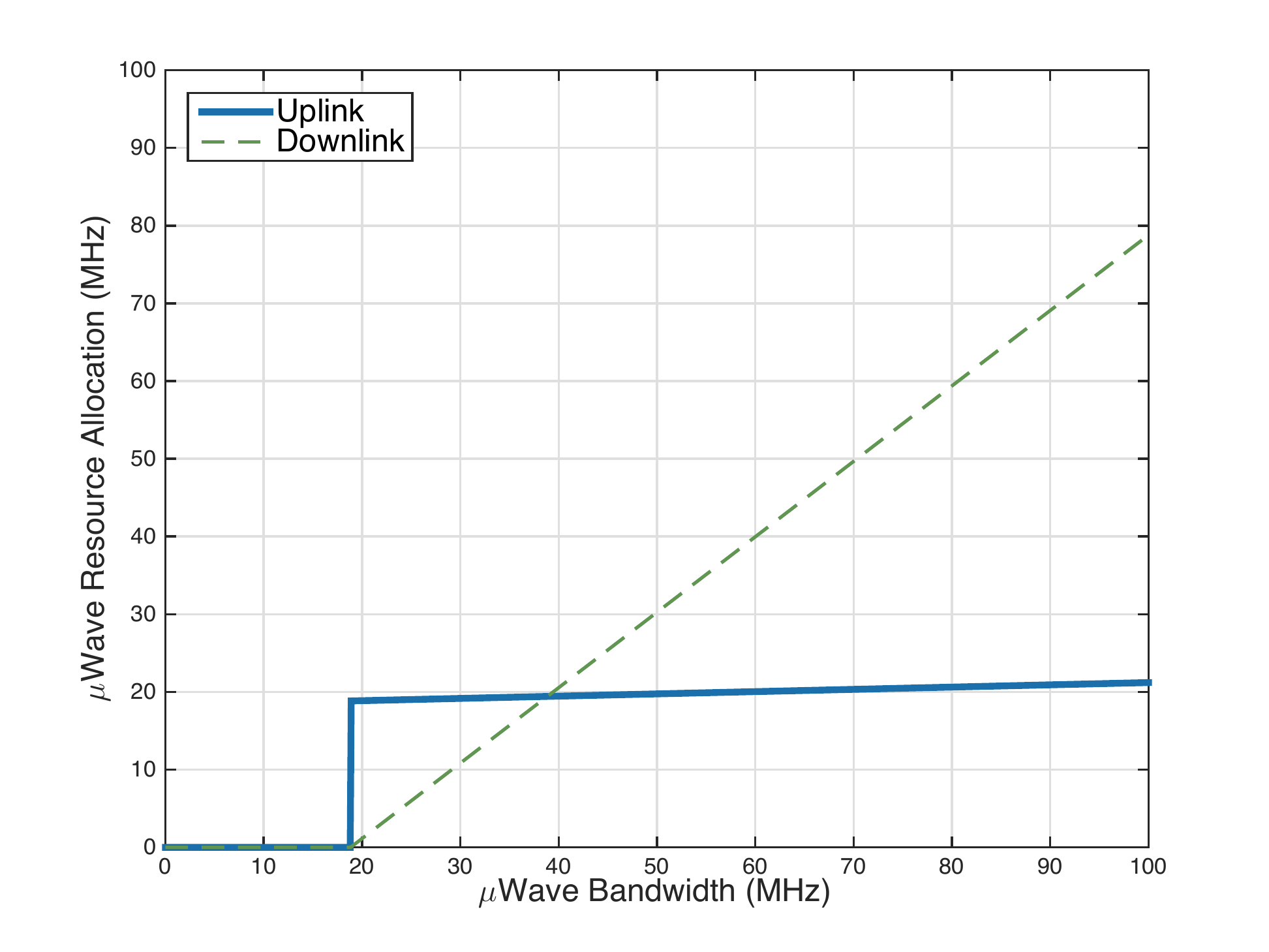} 
	\caption{Uplink and downlink $\mu$Wave resource allocations with mmWave bandwidth $500$ MHz ($\lambda_\text{u} = 0.02$, $\alpha_\mu = 4.58$, $\alpha_\text{m}=5.76$, $\lambda_g = 0.1$, $\theta = 10^\circ$ , $\sigma^2=1$, $T=0.03$).}\label{Fig:Prop3}
\end{figure}

Applying the resource allocation results to \eqref{Eq:RateDL} yields the following uplink rate requirement guaranteeing maximized downlink average rate.
\begin{corollary}\emph{
\emph{(Downlink Average Rate with Minimum Uplink Rate Requirement)} If $W\geq T W_\text{m} \gamma_\text{m}/\gamma_\mu$ for $\lambda_\text{m}, \lambda_\mu \gg \lambda_\text{u}$, maximized downlink average rate while guaranteeing the minimum uplink rate is given as:
\begin{equation} \small
{R_\text{d}}^* = \frac{1}{1 + T}\log\( {1 + c_\text{d}\lambda_\text{m}}^{W_\text{m}\( 1 - \sqrt{\frac{S}{\lambda_\text{m}}} \)\[ \( \frac{\alpha_\text{m}}{2}\)e^{-\lambda_g S} +1 \]} {\lambda_\mu}^{\frac{W \alpha_\mu}{2}} \)
\end{equation}\normalsize
where
\vspace{-8pt}\small\begin{align}
c_\text{d} &= \[ \frac{\pi}{\sigma^2} \l\{ \frac{2 \sigma^2}{\theta} \( \rho_m e^{-\lambda_g S}\lambda_\text{u}  \)^{-\frac{\alpha_\text{m}}{2}} \r\}^{e^{-\lambda_g S}} \]^{W_\text{m}\( 1-\sqrt{\frac{S}{\lambda_\text{m}}}\)}  
\hspace{-10pt} \( \rho_\mu \lambda_\text{u}\)^{-\frac{\alpha_\mu W}{2}}  \nonumber
\end{align} \normalsize 
otherwise, the rate cannot satisfy the uplink rate requirement.
}\end{corollary}
The result reveals BS densification logarithmically increases the downlink rate while $\mu$Wave and mmWave resource amounts and path loss exponents linearly increases the rate. Larger indoor region area under $\lambda_\text{m} \gg S$ exponentially decreases the rate. Additionally, increasing the minimum uplink rate target decreases the downlink rate. This result is visually elucidated by Fig. 6 in Section V.

\subsection{Resource Management and Cell Planning}

This section provides resource management and cell planning guidelines based on the closed-form $\mu$Wave uplink/downlink (see Proposition 1) and mmWave downlink (see Proposition 2) spectral efficiency lower bounds derived in Sections \ref{Sect:muWaveSE} and D. As Fig. 3 and 4 numerically validate the tightness of the lower bounds, we henceforth regard these lower bounds as approximations. To simplify our exposition, we focus on the asymptotic behaviors as $\lambda_\mu, \lambda_\text{m} \rightarrow \infty$.

Consider the uplink requirement feasible condition $W\geq T W_\text{m} \gamma_\text{m}/\gamma_\mu$ from Proposition 3, leading to the minimum number of the required $\mu$Wave BSs along with increasing mmWave BSs.

\begin{proposition}\emph{
\emph{(Required $\mu$Wave BS)} For $\lambda_\mu, \lambda_\text{m} \rightarrow \infty$, guaranteeing the minimum uplink rate requires the following $\mu$Wave and mmWave BS density relation:
\begin{equation}
O(\lambda_\text{m}) = O\({\lambda_\mu}^{\frac{\alpha_\mu W} {T W_\text{m} \[ (\alpha_\text{m}- 2)e^{-\lambda_g S} + 2 \]}}\)
\end{equation}
}\end{proposition}

To achieve the ever-growing downlink average rate while guaranteeing the minimum uplink average rate, the result shows the required $\mu$Wave BS density should be increased with much higher order of the mmWave BS density in practical scenarios where $W_\text{m} \gg W$. The reason is the logarithmic spectral efficiency improvement by $\mu$Wave BS densification (see Proposition 1) cannot overtake the linearly increased mmWave downlink rate. 

Ameliorating this situation in practice therefore requires to procure additional $\mu$Wave resource as Fig. 7 visualizes in Section V. The following Corollary investigates how much amount of $\mu$Wave resource is required to achieve the goal when $\mu$Wave BS densification is linearly proportional to the mmWave BS's.

\begin{corollary}\emph{\emph{(Required $\mu$Wave Resource)} For $\lambda_\mu \rightarrow \infty$ and $\lambda_\text{m} = p \lambda_\mu$ for a constant $p>0$, the minimum uplink average rate constraint requires $\mu$Wave resource amount  as follows.
\small\begin{equation}
W \geq \frac{T W_\text{m}}{\alpha_\mu } \[ (\alpha_\text{m}- 2)e^{-\lambda_g S} + 2 \]
\end{equation}\normalsize
}\end{corollary}

The result insists that even when the number of $\mu$Wave BSs increases with the same order of the mmWave BSs', it is imperative to procure additional $\mu$Wave resource in order to increase downlink average rate with assuring the minimum uplink rate. It is visually elaborated by Fig. 5 in Section V.

%

\begin{figure}
\centering
\includegraphics[width=9cm]{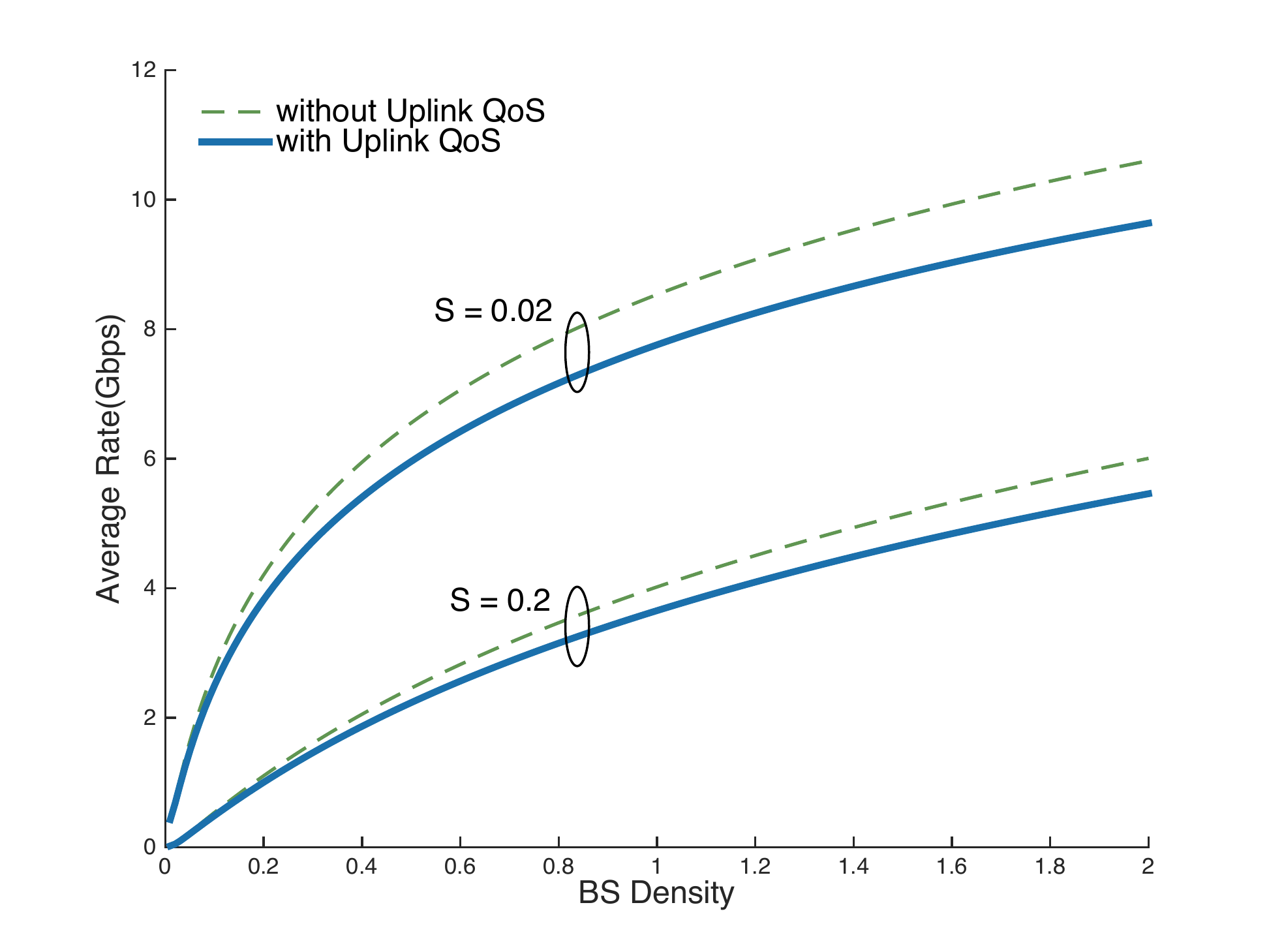}   \vspace{-20pt}
	\caption{Maximized downlink average rate with and without the uplink rate requirement $T=0.1$ under $\mu$Wave bandwidth $20$ MHz and mmWave bandwidth $500$ MHz ($\lambda_\text{u} = 0.02$, $\alpha_\mu = 4.58$, $\alpha_\text{m}=5.76$, $\lambda_g = 0.1$, $\theta = 10^\circ$ , $\sigma^2=1$).}
	\label{Fig:Corollary1}
\end{figure}

\begin{figure}
\centering
\vspace{-12pt}\includegraphics[width=9cm]{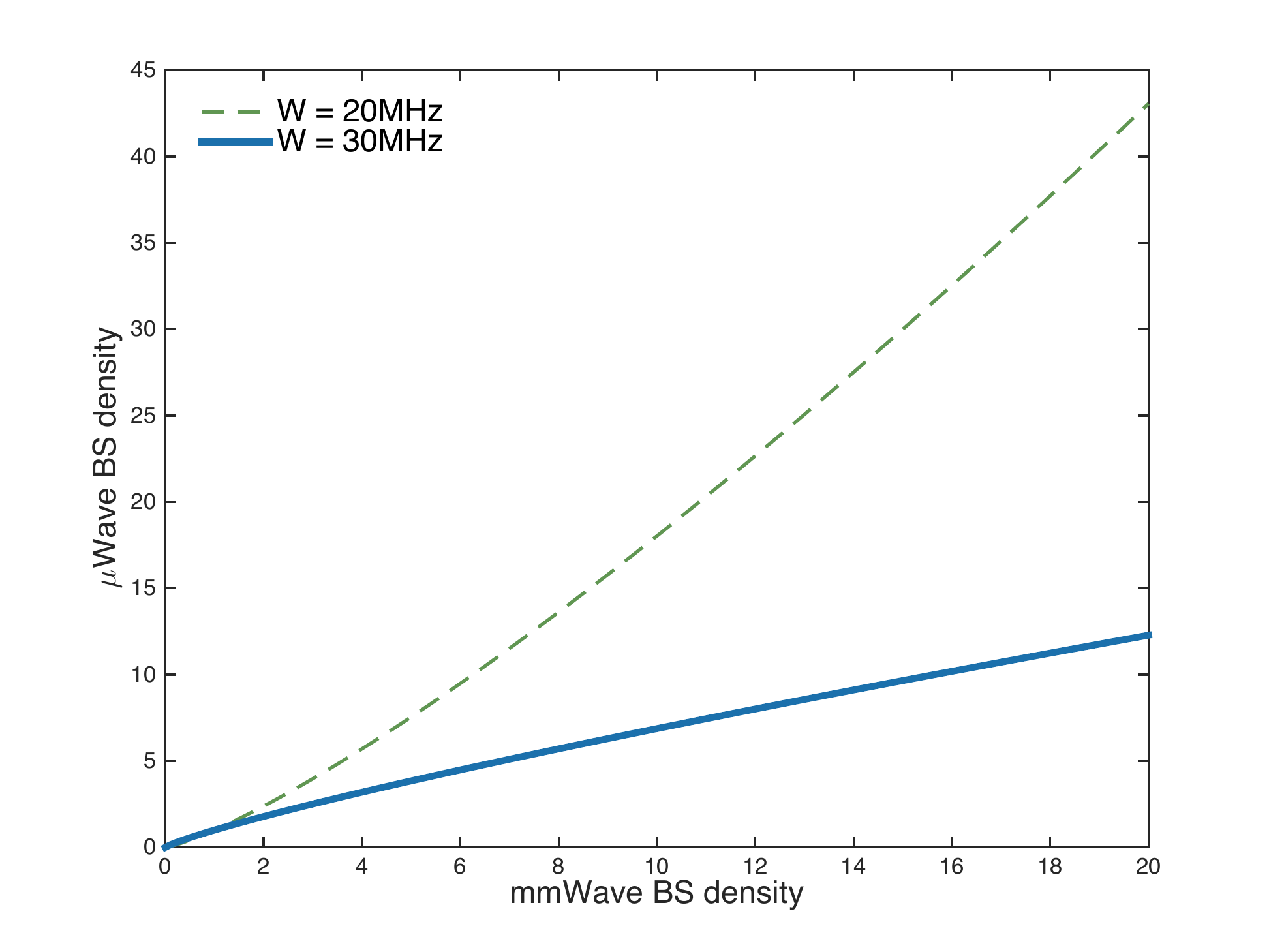}  
	\caption{Minimum required $\mu$Wave BS density as mmWave BS density increases for $\mu$Wave bandwidths $20$ MHz and $30$ MHz under mmWave bandwidth $500$ MHz ($T = 0.04$, $\lambda_\text{u} = 0.02$, $\alpha_\mu = 4.58$, $\alpha_\text{m}=5.76$, $\lambda_g = 0.1$, $S = 0.02$, $\theta = 10^\circ$ , $\sigma^2=1$).}
	\label{Fig:SEmm_out}
\end{figure}

\section{Numerical Results} \label{Sect:NumResult}
This section visualizes the resource management and cell planning guidelines proposed in Section IV under practical scenarios. According to \cite{Rappaport5G:13}, path loss exponents are set as: $\alpha_\mu = 4.58$, $\alpha_\text{m} = 5.76$. The indoor mmWave path loss exponent is set as 2 as assumed in Section \ref{Sect:Channel}. Additionally, user density $\lambda_\text{u}=0.02$, indoor region density $\lambda_g = 0.1$, main lobe beam width $\theta=10^\circ$, and noise power $\sigma^2$ is fixed as $1$. The radio resource amount for $\mu$Wave is set as $20$ MHz as default, and the amount for mmWave as $500$ MHz \cite{SamsungGC:13}.

Fig. 5 illustrates Proposition 3 and Corollary 2 that shows the uplink (thick blue) and downlink (thin green) $\mu$Wave resource allocations. For the given environment, guaranteeing the uplink average rate by $3\%$ of the downlink rate requires at least $19$ MHz uplink $\mu$Wave bandwidth. Considering the current $20$ MHz $\mu$Wave bandwidth, It implies implies even achieving such low uplink/downlink ratio requires to dedicate most of the $\mu$Wave bandwidth to the uplink due to severe uplink/downlink rate asymmetry. This contradicts with the current resource allocation tendency that is likely to allocate more resource to downlink \cite{Rel12Beyond:13}. Focusing on the curve slopes once the minimum uplink rate requirement is achieved, it shows the surplus $\mu$Wave bandwidths are mostly allocated to the downlinks so as to maximize the downlink rates.


Fig. 6 corresponds to Corollary 1 that illustrates the maximized downlink rate with assuring the minimum uplink/downlink rate ratio $T$ along with increasing mmWave BS density $\lambda_\text{m}$. The figure compares the effect of the uplink rate requirement on the resultant downlink rate (solid versus dashed), capturing the downlink rate is pared down to the point achieving the minimum uplink rate. Moreover, the figure revels the effect of the indoor region area $S$ (blue versus green) that larger $S$ leads to the lower downlink rate as expected in the discussion of Corollary 1. In addition, the figure reveals that $1$ Gbps downlink average rate is achieved when mmWave BS density is $1.5$ times larger than the user density for $S=0.02$ and when it is $10$ times larger for $S=0.2$.

Fig. 7 visualizes Proposition 4 that validates the proposed resource management and cell planning guidelines. It shows $\mu$Wave BS densification cannot independently cope with the uplink rate requirement problem, but requires the aid of procuring more $\mu$Wave spectrum. The impact of procuring more $\mu$Wave spectrum is observed by the curve increasing tendencies in the figure. The $\mu$Wave BS density for $\mu$Wave bandwidth $20$ MHz (dotted green) shows power-law increase while the density for $30$ MHz (solid blue) does sub-linear increase. These different required $\mu$Wave BS increasing rates corroborate the necessity of the additional $\mu$Wave spectrum. The figure in addition reveals that the effect of indoor region area $S$ that decreases the minimum required $\mu$Wave BS density due to its downlink rate reduction discussed in Corollary 1 and visualized in Fig. 6.



\section{Conclusion} \label{Sect:Conclusion}

In this paper we propose a mmWave overlaid ultra-dense cellular network operating downlink transmissions via both mmWave and $\mu$Wave bands whereas uplink transmissions only via $\mu$Wave band due to its technical implementation difficulty at mobile users. Regarding this asymmetric uplink and downlink structure, we provide the resource management and cell planning guidelines so as to maximize downlink average rate while guaranteeing the minimum uplink rate (see Propositions 3 and 4 as well as Corollary 2). Such results are calculated on the basis of the closed-form mmWave (see Proposition 1) and $\mu$Wave (see Proposition 2) spectral efficiencies derived by using stochastic geometry. 

The weakness of this study is the use of an arbitrary indoor region (or blockage) area $S$ when calculating mmWave spectral efficiencies, which may alter the network design results. Moreover, the blockage modeling in this paper is necessary to be compared with the recent mmW blockage analysis such as \cite{Heath:13,Kulkarni:14}. Further extension should therefore contemplate more realistic building statistics as well as a rigorous comparison with the preceding works.


%

\bibliographystyle{ieeetr}

\end{document}